\documentclass[a4paper,10pt,prl,twocolumn,showpacs,aps,floats,floatfix,superscriptaddress]{revtex4}

\usepackage{epsfig}
\begin{document}

\newcommand{\avk}{\langle k \rangle}
\newcommand{\fluck}{\langle k^2 \rangle}

\title{Weighted evolving networks: coupling topology and weights
dynamics}


\author{Alain Barrat} \affiliation{Laboratoire de Physique Th{\'e}orique (UMR du CNRS
  8627), Batiment 210, Universit{\'e} de Paris-Sud 91405 Orsay, France}
\author{Marc Barth\'elemy} \affiliation{CEA-Centre d'Etudes de
  Bruy{\`e}res-le-Ch{\^a}tel, D\'epartement de Physique Th\'eorique et
  Appliqu\'ee BP12, 91680 Bruy{\`e}res-Le-Ch{\^a}tel, France}
\author{Alessandro Vespignani}
\affiliation{Laboratoire de Physique Th{\'e}orique (UMR du CNRS 8627),
  Batiment 210, Universit{\'e} de Paris-Sud 91405 Orsay, France}


\widetext
\begin{abstract}

We propose a model for the growth of weighted networks that couples
the establishment of new edges and vertices and the weights' dynamical
evolution.  The model is based on a simple weight-driven dynamics and
generates networks exhibiting the statistical properties observed in
several real-world systems. In particular, the model yields a
non-trivial time evolution of vertices' properties and scale-free
behavior for the weight, strength and degree distributions.

\end{abstract}


\maketitle 


The last years have witnessed a hectic activity devoted to the
characterization and understanding of networked structures as diverse
as ecological and biological systems or the Internet and the
WWW~\cite{Barabasi:2000,Amaral:2000,mdbook,psvbook}.  These networks
generally exhibit complex topological properties such as the
small-world phenomenon~\cite{watts98} and scale-free
behavior~\cite{Barabasi:2000}. The need for explaining these complex
topological features has led to a new modeling framework that focuses
on the dynamical evolution and growth of networks
\cite{Barabasi:1999}. In this perspective, a wide array of models
aimed at capturing various properties of real networks have been
formulated~\cite{mdbook}.  These models, however, do generally
consider only the topological structure and do not take into account
the interaction strength \---the weight of the link\--- that
characterizes real networks. For example, the diversity of
the predator-prey interaction and  of  metabolic reactions 
is considered a critical ingredient
of ecosystems~\cite{Pimm,Krause:2003} 
and metabolic networks~\cite{Almaas:2004}, respectively; in social systems,
the weight of interactions is very important in the characterization
of the corresponding networks~\cite{Granovetter}. 
Similarly, the Internet traffic~\cite{psvbook} or the number of passengers in the
airline network~\cite{Amaral:2000,Guimera:2003,Barrat:2003} are
crucial quantities in the study of these systems. Interestingly,
recent studies~\cite{Almaas:2004,Barrat:2003,Li:2003a,Garla:2003} have
shown that weighted networks exhibit additional complex properties
such as broad distributions and non-trivial correlations of weights
that do not find an explanation just in terms of the underlying
topological structure.

In this Letter, we define a simple model for weighted networks that
considers the basic evolution of the system as driven by the weight
properties of edges and vertices.  In addition, differently from early
models proposed in the past~\cite{Yook:2001,zheng03}, we allow the
dynamical evolution of weights during the growth of the system. 
This mimics the evolution and reinforcements of 
interactions in natural and infrastructure networks.  
The generated networks display power-law behavior for the weight, 
degree and strength distributions with non-trivial exponents 
depending on the unique parameter defining the model's dynamics.
These results suggest that the inclusion of weights in  networks 
modeling naturally explains the diversity of scale-free behavior 
empirically observed in real networked structures. Strikingly, 
the weight-driven growth recovers an effective
preferential attachment for the topological properties, providing a
microscopic explanation for the ubiquitous presence of this mechanism.
  
Weighted networks are usually described by a matrix $w_{ij}$
specifying the weight on the edge connecting the vertices $i$ and $j$,
with $i,j= 1,...,N$ where $N$ is the size of the network ($w_{ij}=0$
if the nodes $i$ and $j$ are not connected). In the following we will
consider only the case of symmetric weights $w_{ij}=w_{ji}$.
Prototypical examples of weighted networks can be found in the
world-wide airport network (WAN)~\cite{Guimera:2003,Barrat:2003} and
the scientific collaboration network (SCN)~\cite{newmancoll,vicsek}.
In the airport network each given weight $w_{ij}$ is the number of
available seats on direct flights connections between the airports $i$
and $j$. In the SCN the nodes are identified with authors and the
weight depends on the number of co-authored
papers~\cite{newmancoll,Barrat:2003}.  A first topological
characterization of networks is obtained with the vertices degree $k$, 
number of connected neighbors, and the associated probability 
distribution $P(k)$. 
Complex networks often exhibits a power-law degree distribution $P(k)\sim
k^{-\gamma}$ with $2\leq\gamma\leq 3$ and measurements of weighted
networks confirm this evidence
~\cite{Guimera:2003,Barrat:2003}. Similarly, a first characterization
of weights is obtained by the distribution $P(w)$ that any given edge
has weight $w$. In complex networks such as the WAN and SCN this
distribution is heavy tailed and spans several orders of
magnitude~\cite{Barrat:2003,Li:2003a}.  Along with the degree of a
node, a very significative measure of the network properties in terms
of the actual weights is obtained by looking at the vertex {\em
strength} $s_i$ defined as~\cite{Yook:2001,Barrat:2003}
\begin{equation}
s_i=\sum_{j\in{\cal V}(i)}w_{ij},
\end{equation}
where the sum runs over the set ${\cal V}(i)$ of neighbors of $i$.
The strength of a node integrates the information about its
connectivity and the weights of its links, and can
be considered as the natural generalization of the connectivity. For
instance, in the case of the WAN the strength provides the actual
traffic going through the vertex $i$ and is an obvious measure of the
size and importance of each airport. This quantity obviously increases
with the vertex degree $k_i$ and usually displays the power-law
behavior $s\sim k^{\beta}$, with the exponent $\beta$ varying as a
function of the specific network. Finally, we note that in most cases, 
the empirical distribution of $s$ is 
heavy-tailed ~\cite{Barrat:2003}, analogously to the power-law 
decay of the degree distribution. 

Previous approaches to the modeling of weighted networks focused on
growing topologies where weights were assigned statically, i.e. once
for ever, with different rules related to the underlying
topology~\cite{Yook:2001,zheng03}. These mechanisms, however, overlook
the dynamical evolution of weights according to the topological
variations. We can illustrate this point in the case of the airline
network. If a new airline connection is created between two airports
it will generally provoke a modification of the existing traffic of
both airports. In general, it will increase the traffic activity
depending on the specific nature of the network and on the local
dynamics.  In the following we will implement a model that takes into
account the coupled evolution in time of topology and weights.

\begin{figure}[t]
\begin{center}
\epsfig{file=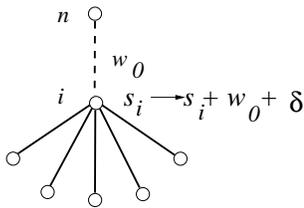,width=4cm}
\end{center}
\caption{ Illustration of the construction rule. A new node $n$
connects to a node $i$ with probability proportional to
$ s_i/\sum_j s_j$. The weight of the new edge is $w_0$ and the
total weight on the existing edges connected to $i$ is modified by an amount equal
to $\delta$. }
\label{fig:rule}
\end{figure}
The model dynamics starts from an initial seed of $N_0$ vertices
connected by links with assigned weight $w_0$.  At each time step, a
new vertex $n$ is added with $m$ edges that are randomly attached to a
previously existing vertex $i$ according to the probability
distribution
\begin{equation}
\Pi_{n\to i}=\frac{s_i}{\sum_j s_j}.
\label{sdrive}
\end{equation}
This rule relaxes the usual degree preferential attachment, focusing
on a strength driven attachment in which new vertices connect more
likely to vertices handling larger weights and which are more central
in terms of the strength of interactions. The weight of each new 
edge is fixed to a value $w_0$. Moreover, the presence of the
new edge $(n,i)$ will introduce variations of the existing weights 
across the network. In particular, we consider the local
rearrangements of weights between $i$ and its neighbors $j\in{\cal
V}(i)$ according to the simple rule
\begin{equation}
w_{ij}\to w_{ij}+\Delta w_{ij},\ \ \ 
\end{equation}
where 
\begin{equation}
\Delta w_{ij}=\delta\frac{w_{ij}}{s_i}.
\label{eq:rule}
\end{equation} 
This rule considers that the establishment of a new 
edge of  weight $w_0$ with the vertex $i$ 
induces a total increase of traffic $\delta$  that is proportionally 
distributed among the edges departing from the vertex according to their
weights (see Fig.~\ref{fig:rule}), yielding $s_i\to s_i+\delta+w_0$.
At this stage, it is worth remarking that while we  will focus on
the simplest model with $\delta=const$, different choices 
of $\Delta w_{ij}$ with  heterogeneous $\delta_i$ or 
depending on the specific properties of  each
vertex ($w_{ij}, k_i, s_i$) can be considered~\cite{Note}.
Finally, after the weights have been updated the growth process is iterated
by introducing a new vertex with the corresponding re-arrangement of weights.

The weight driven attachment (Eq.~(\ref{sdrive})) appears to be a
plausible mechanism in many technological networks. In the Internet
new routers connect to more central routers in terms of bandwidth and
traffic handling capabilities and in the airport networks new
connections are generally established to airports with a large
passenger traffic. Even in the SCN this mechanism might play a role
since an author with more co-authored papers is more visible and open
to further collaborations. Also the weights dynamics can be traced
back to simple dynamical processes. 
Indeed, the  
quantity $w_0$ sets  the scale of the weights and without loss of
generality we can use the rescaled quantities $w_{ij}/w_0$, $s_i/w_0$
and $\delta/w_0$.  For the sake of simplicity we can set $w_0=1$ and 
the model depends only on the dimensionless parameter $\delta$, that is
the fraction of weight which is `induced' by the new edge onto the
others. 
According to the value of $\delta$, different scenarios are
possible. If the induced weight is $\delta\approx w_0 =1$ we mimic
situations in which an appreciable fraction of traffic generated by
the new connection will be dispatched in the already existing
connections. This is plausible in the airport networks where the
transit traffic is rather relevant in hubs.  In the case of $\delta <
1$ we face situations such as the SCN where it is reasonable to
consider that the birth of a new collaboration (co-authorship) is not
triggering a more intense activity on previous collaborations.
Finally, $\delta>1$ is an extreme case in which a new edge generates a
sort of multiplicative effect that is bursting the weight or traffic
on neighbors.

In the model, the time
is measured with respect to the number of vertices added to the graph,
resulting in the definition $t=N-N_0$, and the natural time scale of
the model dynamics is the network size $N$.
The network's evolution
can be inspected analytically by studying the time evolution of the
average value of $s_i(t)$ and $k_i(t)$ of the $i$-th vertex at time
$t$, and by relying on the continuous approximation that treats $k$,
$s$ and the time $t$ as continuous
variables~\cite{Barabasi:2000,mdbook}. When a new edge is added to the
network, the strength $s_i$ of vertex $i$ can increase either if the
edge connects directly to $i$ or to one of its neighbor. The evolution
equation for $s_i$ is thus given by
\begin{eqnarray}
\nonumber
\frac{ds_i}{dt}&=&m\frac{s_i}{\sum_js_j}(1+\delta)+
\sum_{j\in{\cal V}(i)}m\frac{s_j}{\sum_l s_l}
\delta\frac{w_{ij}}{s_j}\\
&=&\frac{2\delta+1}{2\delta+2}\frac{s_i(t)}{t},
\label{eq_evol_s}
\end{eqnarray}
where the last expression is recovered by  noticing that each 
added link increases the total strength by
an amount equal to $2+2\delta$ thus implying that 
$\sum_i s_i(t)\approx 2m(1+\delta)t$.
\begin{figure}[t]
\begin{center}
\epsfig{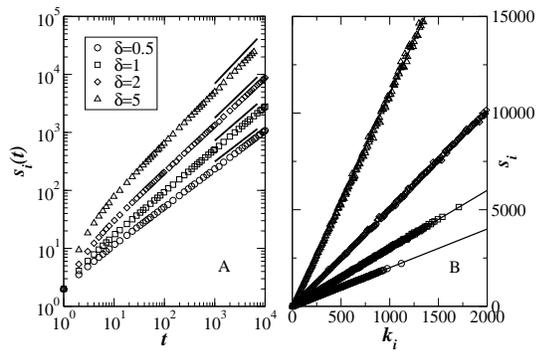}
\end{center}
\caption{ (A) Evolution of the strength of vertices during the growth
of the network, for various values of $\delta$; The thick lines
are the predicted power laws $t^a$, $a=(1+2\delta)/(2+2\delta)$ 
($m=2$, $N=10^4$). (B) Strength $s_i$ versus $k_i$ for various
values of $\delta$; the straight lines are the predictions
$s_i=(1+2\delta)k_i$.
}
\label{fig:st}
\end{figure}
This equation can be readily integrated with the 
initial condition $s_i(t=i)=m$, yielding 
\begin{equation}
s_i(t)=m ~\left(\frac{t}{i}\right)^{\frac{2\delta+1}{2\delta+2}}
\label{eq_s.vs.t}
\end{equation}
Similarly, the rate equation for the degree evolution reads 
\begin{equation}
\frac{dk_i}{dt}=m\frac{s_i(t)}{\sum_js_j(t)}=\frac{s_i(t)}{2(1+\delta)t},
\label{eq_ki}
\end{equation}
that implies $k_i(t)=s_i(t)/(2\delta+1)$.
The proportionality relation $s\sim k$, that implies $\beta=1$, 
is particularly relevant since it states that the 
weight-driven dynamics generates in Eq.~(\ref{sdrive}) 
an effective degree preferential attachment that is parameter independent.
This highlights an alternative microscopic mechanism accounting
for the presence of the preferential attachment dynamics in growing
networks.
\begin{figure}[t]
\begin{center}
\epsfig{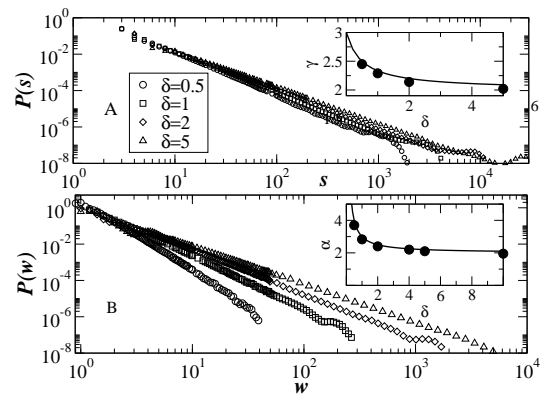}
\end{center}
\caption{(A) Probability distribution $P(s)$. Data fitting points 
(filled circles) are consistent with the analytic prediction of  power-law behavior $s^{-\gamma}$ with $\gamma=(3+4\delta)/(1+2\delta)$ (solid line) as shown in the inset. (B) Probability distribution of the 
weights $P(w)\sim w^{-\alpha}$. In the inset we report the value 
of $\alpha$ obtained by data fitting
(filled circles) and the analytic expression $\alpha=2+1/\delta$ (solid line).
The data are averaged over $200$ networks of size $N=10^4$.
}
\label{fig:P_s}
\end{figure}

The behavior of the various statistical distributions can be computed
by noticing that the time at which the node $i$ enters 
the network is uniformly distributed in [0,t] and 
\begin{equation}
  P(k, t) = \frac{1}{t+N_0} \int_0^{t} \delta(k - k_i(t)) di, 
\label{eqdistr}
\end{equation}
where $\delta(x)$ is the Dirac delta function. Solving the above
integral and considering the infinite size limit $t\to\infty$  yields 
$P(k)\sim k^{-\gamma}$ with
\begin{equation}
\gamma=\frac{4\delta+3}{2\delta+1}.
\end{equation}
It is straightforward to obtain the same behavior $P(s)\sim
s^{-\gamma}$ for the strength distribution since $s$ and $k$ are
proportional.  This result shows that the obtained graph is a
scale-free network described by an exponent $\gamma\in[2,3]$ that
depends on the value of the parameter $\delta$.  In particular, when
the addition of a new edge doesn't affect the existing weights
($\delta=0$), the model is topologically equivalent to the
Barabasi-Albert model \cite{Barabasi:1999} and the value $\gamma=3$ is
recovered.  For larger values of $\delta$ the distribution is
progressively broader with $\gamma\to 2 $ when $\delta\to \infty$.
This indicates that the weight-driven growth generates scale-free
networks with exponents varying in the range of values usually
observed in the empirical analysis of networked
structures~\cite{Barabasi:2000,mdbook,psvbook}. Noticeably the
exponents are non-universal and depend only on the parameter $\delta$
governing the microscopic dynamics of weights. The model therefore
proposes a general mechanism for the occurrence of varying power-law
behaviors without resorting on more complicate topological rules and
variations of the basic preferential attachment mechanism.

In order to check the analytical predictions we performed numerical
simulations of networks generated by using the present model with 
different value of $\delta$, minimum degree $m$ and varying size $N$.
In Fig.~\ref{fig:st} we show the behavior of the vertices' strength versus
time for different value of $\delta$, recovering the behavior predicted analytically. 
We also report the average strength  $s_i$ of vertices 
with degree $k_i$  and confirm that $\beta=1$ as well as the
value of the pre-factor $2\delta+1$. Finally, in  Fig.~\ref{fig:P_s}A 
we plot the probability distribution $P(s)$ that exhibits a power-law behavior 
in excellent  agreement with the theoretical predictions.

\begin{figure}[t]
\begin{center}
\epsfig{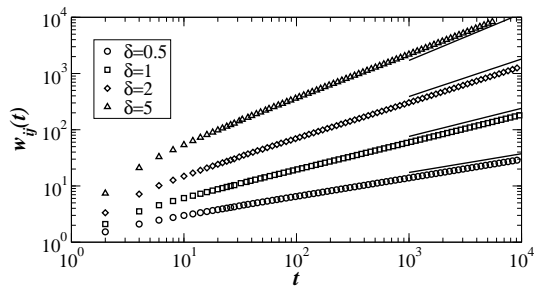}
\end{center}
\caption{Time evolution of $w_{ij}$ during the
growth of the network ($m=2$ and
$N=10^4$) for different values of $\delta$.
The functional behavior is consistent with the predicted power laws $t^b$,
$b=\delta/(1+\delta)$, shown by thick lines.
}
\label{fig:w}
\end{figure}

Similarly to the previous quantities,  it is possible to 
obtain analytical  expressions for the evolution of  weights 
and the relative statistical distribution.  
The weight $w_{ij}$ increases by the addition of a new link 
either on $i$ or on $j$ and the corresponding
rate  equation can  be written as 
\begin{equation}
\frac{dw_{ij}}{dt}=m\frac{s_i}{\sum_js_j}\delta\frac{w_{ij}}{s_i}
+m\frac{s_j}{\sum_js_j}\delta\frac{w_{ij}}{s_j}.
\end{equation}
This equation can be solved with the initial condition
$w_{ij}(t_{ij})=1$, where $t_{ij}=$max$(i,j)$ is the time at which the
edge is established, yielding 
$w_{ij}(t)=(t/t_{ij})^{\delta/(\delta+1)}$
From this expression it is possible to calculate similarly to $P(k)$
the probability distribution $P(w)$ which is in this case also a
power-law $P(w)\sim w^{-\alpha}$ where 
\begin{equation}
\alpha=2+\frac{1}{\delta}.
\end{equation}
The exponent $\alpha$ has large variations as a function of the parameter
$\delta$ and $P(w)$ moves from a delta function for $\delta=0$ to a
very slow decaying power-law with $\alpha=2$ if
$\delta\to\infty$. This feature clearly shows that the weight
distribution is extremely sensitive to changes in the microscopic
dynamics ruling the network's growth.  These predictions can be
confirmed by numerical simulations and the results are shown in
Figs.~\ref{fig:P_s}B and \ref{fig:w}.

The model we have presented is possibly the simplest one in the class
of weight-driven growing networks.  A novel feature in the model is
the weight dynamical evolution occurring when new vertices and edges
are introduced in the system. This simple mechanism produces a wide
variety of complex and scale-free behavior depending on the physical
parameter $\delta$ that controls the local microscopic dynamics. While
a constant parameter $\delta$ is enough to produce a wealth of
interesting network properties, a natural generalization of the model
consists in considering $\delta$ as a function of the vertices degree
or strength. For example, a non-linear generalization of eq. (\ref{eq:rule})
can yield non-linear correlations between strengths and degrees
($\beta\neq 1$)~\cite{Note}.  Similarly, more complicated variations of
the microscopic rules may be implemented to mimic in a detailed
fashion particular networked systems. In this perspective the present
model appears as a general starting point for the realistic modeling
of complex weighted networks.
\begin{acknowledgments}
We thank R. Pastor-Satorras and M. Vergassola for useful
discussions.
A.B and A. V. are partially funded by the 
EC-Fet  project COSIN IST-2001-33555.
\end{acknowledgments}





\vskip -.6cm 

\begin{thebibliography}{99}


\bibitem{Barabasi:2000} 
R.~Albert and  A.-L.~Barab{\'a}si,
Rev. Mod. Phys. {\bf 74}, 47 (2000).

\bibitem{Amaral:2000} L.A.N.~Amaral, A.~Scala, M.~Barth\'elemy, and
H.E.~Stanley, Proc. Natl. Acad. Sci. (USA) {\bf 97}, 11149 (2000).

\bibitem{mdbook}
S.~N. Dorogovtsev and J. F.~F. Mendes, 
{\em Evolution of networks: From biological nets to the
{I}nternet and {WWW}} (Oxford University Press, Oxford, 2003).

\bibitem{psvbook}
R.~Pastor-Satorras and A.~Vespignani,
{\em Evolution and structure of the Internet: A statistical physics 
approach} (Cambridge University Press, Cambridge, 2004).

\bibitem{watts98}
D.J. Watts and S.H. Strogatz,
Nature {\bf 393}, 440 (1998).

\bibitem{Barabasi:1999} 
A.-L.~Barab\'asi and R.~Albert, Science {\bf 286}, 509 (1999)


\bibitem{Pimm} S.L.~Pimm, {\it Food Webs}, The University of Chicago
Press (2nd edition, 2002).

\bibitem{Krause:2003} A.E.~Krause, K.A. Frank, D.M. Mason, R.E. Ulanowicz, 
and W.W. Taylor, Nature {\bf 426}, 282 (2003).

\bibitem{Almaas:2004} E. Almaas, B. Kov\'acs, T. Viscek, Z.N. Oltval and
A.L.~Barab\'asi, Nature {\bf 427}, 839 (2004).

\bibitem{Granovetter} M.~Granovetter, American Journal of Sociology,
{\bf 78} (6) 1360 (1973).


\bibitem{Guimera:2003} R.~Guimera, S.~Mossa, A.~Turtschi, and
L.A.N.~Amaral, Preprint cond-mat/0312535 (2003).

\bibitem{Barrat:2003} A.~Barrat, M.~Barth\'elemy, R.~Pastor-Satorras,
and A.~Vespignani, Proc. Natl. Acad. Sci. USA {\bf 101}, 3747 (2004).

\bibitem{Li:2003a} C.~Li and G.~Chen, cond-mat/0309236,cond-mat/0311333.


\bibitem{Garla:2003} D.~Garlaschelli, S. Battiston, M. Castri,
V.D.P. Servedio, and G. Caldarelli, Preprint  cond-mat/0310503 (2003).

\bibitem{Yook:2001} S.H.~Yook, H.~Jeong, A.-L.~Barabasi, and Y. Tu,
Phys. Rev. Lett. {\bf 86}, 5835 (2001).

\bibitem{zheng03}
D. Zheng, S. Trimper, B. Zheng and P.M. Hui,
Phys. Rev. E {\bf 67}, 040102 (2003).

\bibitem{newmancoll}
M. E. J. Newman, Phys. Rev. E {\bf 64}, 016131 (2001);
Phys. Rev. E {\bf 64}, 016132 (2001).

\bibitem{vicsek}
A.-L. Barab\'asi, H. Jeong, Z. Neda, E. Ravasz, A. Schubert, and T. Vicsek,
Physica A {\bf 311}, 590 (2002).

\bibitem{Note} 
A.~Barrat, M.~Barth\'elemy, and A.~Vespignani, in preparation (2004).

\end{thebibliography}
\end{document}